\documentstyle[aps,epsf]{revtex}
\begin{document}
\baselineskip 14pt

\title{Cosmic Neutrinos and their Detection}
\author{C. Hagmann}
\address{Lawrence Livermore National Laboratory \\
7000 East Avenue, Livermore, CA 94550}

\maketitle

\begin{abstract}

The standard Big-Bang theory predicts a cosmic neutrino background 
with an average number density of $\sim 100/{\rm cm^3}$ per flavor.
The most promising way of its detection is measuring 
the feeble ``neutrino wind'' forces exerted on macroscopic targets. 
The expected acceleration is $\sim 10^{-23} {\rm cm/s^2}$ for Dirac neutrinos  
with a local number density $\sim 10^7/{\rm cm}^3$.
A novel torsion balance design 
is presented, which addresses the sensitivity-limiting factors of 
existing balances, such as seismic and thermal noise, and angular 
readout resolution and stability. 

\end{abstract}

\section{Introduction}

According to the
standard Big Bang model, a cosmic neutrino background, similar
to the familiar microwave background, should exist in our universe.
Because of their weak interactions, cosmic neutrinos fell out
of thermal equilibrium at $t\simeq 1\, \rm sec$, and have been
redshifting since then. Neutrinos of mass $\ll 10^{-3}$ eV 
would still be relativistic today,
have a Fermi-Dirac spectrum with $T\simeq 1.9$ K, and
be spatially uniform with $n_{\nu_L}=n_{\bar\nu_R}\simeq 50/{\rm cm}^3$ per flavor.
In contrast,
neutrinos of mass $\gg 10^{-3}$ eV would be nonrelativistic today, and 
be clustered around galaxies with a typical velocity $v_\nu\simeq 10^{-3}$.
They
contribute to the cosmological energy density an amount 
$\Omega_\nu \simeq \sum_{i} m_{\nu_i} /(92 h^2  {\rm eV})$, where
$h \simeq 0.65 $ is the Hubble expansion rate 
in units of $100\:{\rm km\, s^{-1} Mpc^{-1}}$,
 and where the sum includes all neutrino flavors
that are nonrelativistic today.
Massive neutrinos are a natural 
candidate for the hot component in currently favored
``Mixed Hot+Cold Dark Matter (HCDM)'' \cite{primack98} 
models of galaxy formation. In this scenario, neutrinos would 
contribute $\simeq$ 20 \% ($\sum_{i}m_{\nu_i}\simeq 8\,{\rm eV} $), 
and CDM ({\it e.g.} Wimps and axions) the remainder of 
the dark matter. 
 Another popular model involving a cosmological
constant is $\Lambda\rm HCDM$ \cite{gross98}
with $\Omega_\nu\simeq0.05-0.1$.
The available phase space for massive neutrinos
\cite{tremaine79,ellis94} restricts the local neutrino 
number density to
\begin{equation} 
n_{\nu} <  \,2\times 10^6\, {\rm cm^{-3}} \left(\frac{v_m}{10^{-3}}\right)^3 
\,\sum_{i} \left(\frac{m_{\nu_i}} {10 \rm \,eV}\right)^3. 
\end{equation}  
where $v_m$ is the maximum neutrino velocity in the halo. 
Hence, neutrinos of $m_\nu \sim$ eV, cannot
account for all of the local halo matter density $\sim 300\, \rm MeV/cm^3$.

The detection of cosmic neutrinos by interactions with single electrons or
nucleons is unlikely because of the extremely small rates
and energy deposits.
Past proposals have therefore focused on detecting the coherent mechanical 
effects on macroscopic targets due to the ``neutrino wind''.
In 1974, Stodolsky \cite{stod74} suggested to use 
the energy splitting of an electron,
\begin{equation}
\Delta E= \sqrt{2}\,G_F\,v_e
\left((n_{\nu_e}-n_{\bar\nu_e})-(n_{\nu_\mu}-n_{\bar\nu_\mu})
-(n_{\nu_\tau}-n_{\bar\nu_\tau})\right)
\end{equation}
moving through the neutrino background with velocity $v_e$. However
even for large neutrino number asymmetries, {\it e.g.} $n_{\bar\nu_e}=0$, 
$n_{\nu_{\mu,\tau}}=n_{\bar\nu_{\mu,\tau}}$,
$\Delta E\sim 1.3\times10^{-33} {\rm eV} (v_e/10^{-3}) (n_{\nu_e}/10^7{\rm cm^{-3}})$
is still tiny. In principle, this effect could be measured by observing a torque on a 
permanent magnet, which is shielded against
magnetic noise with superconductors. A benefit of this detection method is that
it works equally well for Dirac and Majorana neutrinos.
Other forces $\propto G_F$ arising from
the reflection or refraction of neutrinos by macroscopic 
objects have been proposed by
several authors \cite{opher74,lewis80,opher82}.
However, all of these ideas
were later found to be flawed \cite{cabi82,lang83}.    

Another approach is to consider forces $\propto G^2_F$ due to random elastic
neutrino scattering \cite{shvarts82,smith83,ferr95,spea96}.    
Here, spatial coherence dramatically increases the cross 
section of targets smaller than the neutrino wavelength 
$\lambda_\nu$. 
In the nonrelativistic limit, one must distinguish between Majorana and Dirac 
neutrinos. For Dirac $\mu$ or $\tau$ neutrinos,
the cross section is dominated by the vector
neutral current contribution
\begin{equation}
\sigma_ D\simeq \frac {G_F^2 m_\nu^2} {8\pi} N_n^2 = 2\times 10^{-55} 
\left(\frac {m_\nu} {10 \rm \,eV}\right)^2 N_n^2\;\;{\rm cm^2}
\end{equation}
where $N_n$ is the number of neutrons in the target
of size
$< \lambda_\nu/2\pi= 20 \,\mu{\rm m}\,(10{\rm\,eV}/m_\nu)(10^{-3}/v_\nu)$. 
For Majorana neutrinos the 
vector contribution to the cross section is suppressed
by a factor $(v_\nu/c)^2$ and potentially the largest contribution arises from the axial
current. The cross section of a spin-polarized target is
$\sigma_M \propto  N_s^2$ ,
where $N_s$ is the number of aligned spins in the grain.

Due to the Sun's peculiar motion $(v_s \sim 220 \rm \,km/s)$ through the galaxy,
a test body will experiences a neutrino wind force
through random neutrino scattering events.   
The wind direction is modulated with the sidereal period of ~23hrs+56min due to
the combined effects of the Earth's diurnal rotation and
annual revolution around the Sun.
For Dirac neutrinos, the acceleration of a target of 
density $\rho$ and optimal radius $ \lambda_\nu/2\pi $ 
has an amplitude \cite{shvarts82,smith83}
\begin{equation}
a= 8\times 10^{-24} \left( \frac {A-Z}{A}\right)^2 \left( \frac 
{v_{\rm s}}{10^{-3}}\right)^2
\left( \frac {n_\nu}{10^7 {\rm cm^{-3}}}\right)
\left( \frac {\rho}{20\,{\rm gcm^{-3}}}\right)
\;{\rm \frac {cm}{s^2}} 
\end{equation}
and is independent of $m_\nu$. For clustered Majorana neutrinos, 
the acceleration is suppressed by a 
factor $\sim 10^6(10^3)$ for an unpolarized (polarized) target.
For unclustered relativistic neutrinos, the force should be aligned with the direction
of the microwave background dipole. Neutrinos of Dirac or Majorana type would 
have the same cross section, giving rise to  
an acceleration
 $\sim 10^{-34}  \rm cm/s^2$.
  A target size much larger than $ \lambda_\nu $ can be assembled, 
while avoiding destructive interference, by using foam-like \cite{shvarts82}
or laminated materials \cite{smith83}.
Alternatively, grains of size  $ \sim \lambda_\nu $ could be randomly embedded
(with spacing $\sim \lambda_\nu $) in a low density host material.

\section{Detector Concept}

The measurement of the neutrino-induced forces and torques will
require major improvements in sensor technology. 
At present, the most sensitive detector of small forces is the 
``Cavendish-type'' torsion balance (see Figure 1), 
which has been widely used for measurements of the gravitational constant 
and searches for new forces. A typical arrangement consists of a  
dumbbell shaped test mass suspended by a 
tungsten fiber inside a vacuum chamber at room temperature. 
The angular deflection is usually read out with
an optical lever. The torsional oscillation frequency is constrained
by the yield strength of the fiber to $>$ 1mHz. Internal friction
in the fiber limits the $Q$ to $< 10^6$. Nonlinearities in the
suspension make the balance quite sensitive
to seismic noise. 
Additional background forces arise from thermal noise and time
varying gravity gradients due to tides, weather, traffic, people etc.
The angular resolution of optical levers is $\sim 10^{-10}
\rm rad$ and the smallest measurable acceleration is $\sim 10^{-13} \rm cm/s^2$
\cite{adel99}.   

%\begin{figure}[ht]      % in second brace, h=here, t=top, b=bottom      
%\centerline{\epsfxsize 5 truein \epsfbox{abazajian0214fig1.eps}}   

\begin{figure} []
\vspace{5mm}
\epsfysize=2.3in
\centerline{\epsfbox{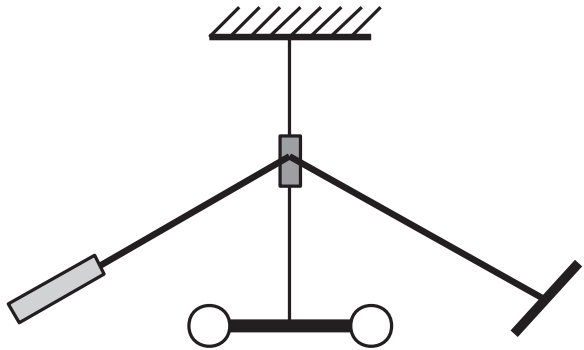}}
\vspace{10 mm}
\caption{Schematic diagram of a ``Cavendish'' torsion balance, which
has been widely used in experimental gravity.} 
\end{figure}

\begin{figure} []
\epsfxsize=3.5in
\centerline{\epsfbox{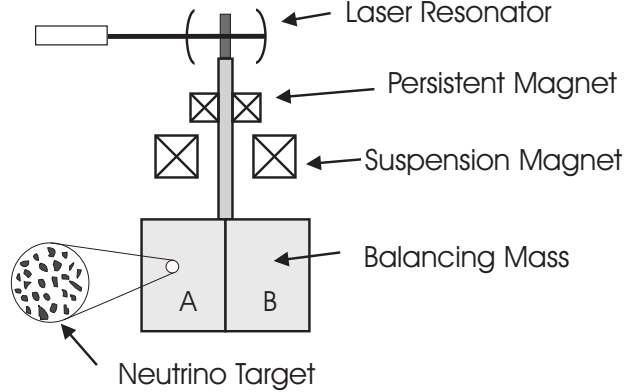}}
\vspace{10 mm}
\caption{Schematic diagram of the torsion oscillator. The target consists
of two hemicylindrical masses with similar densities but different
neutrino cross section. The target of mass $\sim$ kg is suspended 
by a ``magnetic hook''
consisting of a superconducting magnet in persistent mode
floating above a stationary magnet. The rotation angle is read out with
a tunable optical cavity and an ultra-stable laser.} 

\end{figure}

Several improvements seem possible (see Figure 2):
Thermal noise can be decreased by lowering the temperature and by
employing a low dissipation (high $Q$) suspension,
as seen from the expression for the effective thermal 
noise acceleration\cite{brag77book}  
\begin{equation}
a_{\rm th}\simeq 2\times 10^{-23} \left( \frac {T} {1\rm K}\right)^{1/2}
					\left( \frac {\rm 1 kg} {m}\right)^{1/2}
                              \left( \frac {\rm 1 day} {\tau_0}\right)^{1/2}
                              \left( \frac {10^6 \rm s} {\tau}\right)^{1/2}
                              \left( \frac {10^{16}} {Q}\right)^{1/2}
             \; {\rm \frac {cm}{s^2} }
\end{equation}
where $m$ is the target mass, 
$\tau$ is the measurement time, $\tau_0$ is the oscillator period, 
and $T$ is the operating temperature.
A promising high-$Q$
suspension method is a Meissner suspension consisting of a superconducting
body floating above a superconducting coil.
Niobium or NbTi in bulk or film form have been used in the past in such
diverse applications as gravimeters\cite{prot68}, gyros\cite{hard61}, 
and gravitational wave antennas\cite{boughn77,blair79}.
Generally the magnetic field applied to the superconductor is limited
to $B \lesssim 0.2\rm T$ to avoid flux penetration or loss of
superconductivity\cite{boughn77,blair79}. 
However, even for small fields the magnetic flux exclusion is usually 
incomplete in polycrystalline Nb superconductors and 
flux creep will cause noise \cite{karen90}.
The lifting pressure $\propto B^2$ is limited to about $ 100\rm \,g/cm^2$.
This could be dramatically increased by  
replacing the bulk superconductor with a
persistent-mode superconducting magnet, 
as shown in Figure 2. Commonly used NbTi wire has a critical field
of several T. Moreover, the flux lines are strongly pinned by artificial
lattice defects in the magnet wire, 
leading to low levels of flux noise and dissipation.
Very low resistance $\sim 10^{-14}\Omega$ \cite{leup76} 
wire joints can be made by cold welding, and
the magnetic field decay rate can be as low
as $\dot B/B = R/L \sim 10^{-9}/\rm day$, where 
$R$ is the joint resistance and $L$ is the inductance. 
The current decay due to losses in the joint will cause the suspended magnet
to sink gradually.
The concomitant change of magnetization will produce a
slowly varying torque on the float
due to the gyromagnetic (``Einstein-deHaas'') effect.
Due to unavoidable deviations from cylindrical symmetry,
the oscillator will have a very long, but finite rotational oscillation period.
For small amplitudes, the oscillating supercurrents in the wire 
will cause the fluxoids to
move elastically about their pinning centers, and 
very little dissipation is expected.
In order to match the rotation period to the signal period $(\sim 1 \rm\, day)$,
an additional restoring potential may be added. One possibility
is a superconducting or low-loss dielectric ({\it e.g.} single-crystal sapphire)
ellipsoid in the constant electric field of
a parallel plate capacitor\cite{brag77book}.
Both negative and positive torsion coefficients
can be produced by aligning the long axis of the
ellipsoid perpendicular or parallel to the field.     

\begin{figure} [t]
\epsfxsize=3.5in
\centerline{\epsfbox{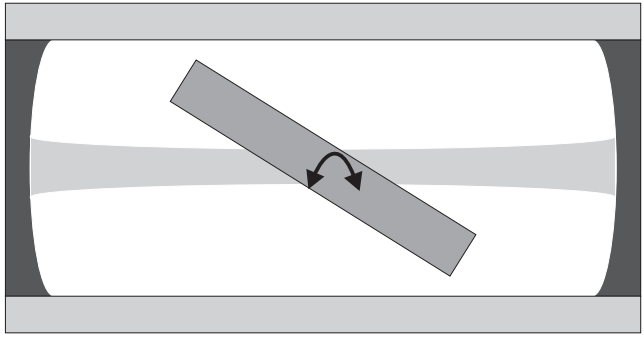}}
\vspace{10 mm}
\caption{Topview diagram of the high-$Q$ tunable optical cavity. 
The cavity is operated in the $\rm TEM_{00p}$
Gaussian mode.
A narrow linewidth $\sim 100 \rm kHz$
is achieved with high relectivity mirrors and
a low-loss dielectric Brewster-angled plate. 
Two symmetrically placed and oppositely oriented
fixed plates may be added to avoid transverse displacement of the beam.} 
\end{figure}

The effects of time-varying gravity gradients can be
reduced with a highly cylindrically symmetric target (see Figure~1).   
With the c.m. of the target centered below the suspension support, 
the leading order gravitational torque arise from the dipole
moment of the torsion balance.
For a source mass $M$ at a distance $R$ from
the balance c.m., the torque is
\begin{equation}
\tau_z\simeq \frac {G_N\,M}{R^3}\, (I_x-I_y)
\end{equation}
where $I_x,I_y$ are the moments of inertia
about the two horizontal axes.
Diurnally varying mass distributions, ${\it e.g.}$ changes
of the atmospheric air mass due to solar heating,
are especially worrisome.
These effects can be minimized with a careful mechanical
design and subsequent balancing steps using a 
movable laboratory source mass.
In addition, locating the experiment in a deep mine
would be beneficial.   
The gravitational interaction of the balance with the Sun and the Moon
are less dangerous, because they
produce torques with a 12 hr period to first order.

Low frequency seismic noise can also be a performance limiting factor in
torsion balance experiments. At a period of $\sim$ 24 hrs, the noise is
dominated by tidal effects, with typical variations
in the strain of $\Delta L/L\sim 10^{-7}$ , in the tilt
about the local (g-referenced) vertical axis of $\Delta\theta\sim 10^{-6}$ rad, 
and in the local gravitational acceleration of $\Delta g/g\sim 10^{-7}$.
Small deviations from ideal symmetry will couple these motions into
the rotational mode of the balance. For example, the tilt-rotation coupling
of a tungsten fiber suspension is typically $\simeq 0.02$\cite{su94}.
Two possible remedies are (1) improving the symmetry of the balance, and
(2) employing active anti-seismic isolation systems. 
A remaining worry is low frequency rotational seismic noise for which no
reliable data is available. It will directly mask the signal and needs to be
compensated. 

The proposed angular rotation readout has very high immunity
against vertical, horizontal, and tilt seismic
noise. It consists of a parametric transducer, which converts the angle to
an optical frequency. As shown in Figure 3, the transducer
consists of a high-$Q$ optical cavity of length $l$, tuned by a Brewster-angled low loss
dielectric plate of thickness $d$. Using high-reflectivity mirrors and a cavity length
$\sim 10\rm\, cm$ should give linewidths of order 100 kHz for the
Gaussian ${\rm TEM}_{00p}$ modes, with
a frequency tuning sensitivity of $df/d\theta \sim f(d/l)$ 
$\sim 10^{14} \rm Hz/rad$. The angular measurement 
precision depends on the number of photons $N$ via 
$\Delta\theta\sim \lambda/(dF\sqrt{N})$, where $F$ is the cavity finesse, and
$\lambda$ is the laser wavelength. This is a factor $\sim F$ better than
the optical lever readout for equal laser power. 
Cryogenic optical resonators have excellent long-term stability and have been
proposed as secondary frequency standards. The measured frequency drifts 
range from
$\sim 1\rm Hz$ over minutes to $\sim 100\rm\, Hz$ over days\cite{seel97}. 
For the measurement of the rotation angle, a stable reference frequency will be
required. This can be easily implemented using a laser locked to a second cavity.  
The described angle readout has little sensitivity to lateral, tilt, and vertical
motion, but couples to rotational noise. A possible solution would be to 
suspend the target as well as the optical cavity in order to suppress the
common rotation mode. The cavity suspension should have a much longer natural period
than the target suspension to avoid its excitation at the signal frequency.
A different approach to suppress rotational noise would employ two identical
torsion balances, but rotated by 180 degrees with respect to each other.

Additional background forces will arise from gas collisions, cosmic ray hits, 
radioactivity, solar neutrino and WIMP interactions
resulting in a Brownian motion of the target.  
The equivalent acceleration is $a\sim (\bar p/m)\sqrt{n/\tau}$, where $\bar p$ is
the average momentum transfer, and $n$ is the collision or decay rate. 
The requirements on the residual gas pressures are 
very severe with only a few collisions with the target 
allowed per second. Cryopumping and the use of getters will be essential.
The cosmic muon flux
at sea level is about $0.01 \rm\, cm^{-2}s^{-1}$ with
a mean energy of $\sim 1 \rm \,GeV$.
Thus, a target of area $100\rm\, cm^2$ 
will experience a collision rate 
$\sim {\cal O}(1) \rm\, s^{-1}$, causing an acceleration
comparable to the signal in Equation 4.  Cosmic
rays can also lead to a net charge buildup on the torsion
balance and spurious electrostatic forces.    
This background can be reduced by several orders of magnitude by
going underground.  
Further disturbing forces may be caused by
time-varying electric and magnetic background fields, which can
be shielded with superconducters. Blackbody radiation and
radiometric effects would be
greatly reduced in a temperature controlled cryogenic
environment. 

Finally, there is a fundamental limit imposed by the uncertainty 
principle. The accuracy of a force measurement obtainable
by continously recording the test mass position (standard quantum limit) is 
\begin{equation}
a_{\rm SQL} = 5\times 10^{-24} \left(\frac{10 {\rm kg}}{ m}\right)^{1/2}
 \left(\frac{\rm 1 day}{\tau_0}  \right)^{1/2}
 \left(\frac{10^6 \rm s}{\tau} \right)\;\;
 \rm \frac{cm}{s^2}
\end{equation}
where $m$ is the test mass, $\tau_0$ is the oscillator period,
and $\tau$ is the measurement time. 
In our proposed position readout, the disturbing back action force arises from
spatial fluctuations in the photon flux passing through the central tuning plate.
A laser beam pulse containing $N$ photons, and centered on the plate rotation axis, 
will produce a random change in the angular momentum of the plate
of $\Delta L\sim F\sqrt{N}\hbar\omega d/c$. 
Here, $d$ is the 
average transverse photon displacement with respect to the rotation axis, 
which is also of order the plate thickness.
A slightly higher sensitivity can be obtained 
with a stroboscopic ``quantum nondemolition'' (QND)
measurement \cite{brag80,brag92book}, 
where the test mass position is recorded every half period.
Here, 
\begin{equation}
a_{\rm QND} = a_{\rm SQL} \, \sqrt{\omega_0 \Delta t}
\end{equation}
and $\Delta t$ is the strobe duration.

There is hope that cosmic neutrinos will be detected in the laboratory early in
the next century, especially if they have masses in the eV range and are of 
Dirac type. The most viable way seems to be a greatly improved torsion balance
operated in a deep mine.

Naturally, the proposed torsion balance could also be applied 
to tests of the weak equivalence principle with the Sun as
the source mass\cite{roll64,brag72}, and to searches for
new short range $(\sim 1 \rm\, m)$ forces 
with a movable laboratory mass\cite{mood84,adel91,su94}.

\acknowledgements
This research was performed by LLNL under the auspices of the U.S. 
Department of Energy under
contract no. W-7405-ENG-48.
I thank E. Adelberger, S. Baessler, I. Ferreras, A. Melissinos,
R. Newman, P. Smith, and W. Stoeffl for useful discussions.

\end{document}